\documentclass[pdftex,twocolumn,epjc3]{svjour3}

\RequirePackage[T1]{fontenc}
\usepackage{amsmath}
\RequirePackage{graphicx}
\RequirePackage{mathptmx}      
\RequirePackage{flushend}
\RequirePackage[numbers,sort&compress]{natbib}
\RequirePackage[colorlinks,citecolor=blue,urlcolor=blue,linkcolor=blue]{hyperref}

\usepackage{color}
\definecolor{darkgreen}{rgb}{0,0.5,0}
\definecolor{darkblue}{rgb}{0,0,0.7}
\definecolor{darkred}{rgb}{0.5,0,0.0}
\definecolor{darkorange}{rgb}{0.8,0.4,0.0}

\usepackage{amsmath}
\usepackage{amssymb}
\usepackage{xspace}
\usepackage{subfig}

\journalname{Eur. Phys. J. C}

\newcommand{\ptj}{p_{t\,J}}
\newcommand{\ptZ}{p_{t\, Z}}
\newcommand{\zcut}{z_{\text{cut}}}
\newcommand{\zc}{z_{\text{cut}}}
\newcommand{\SD}{SoftDrop\xspace}

\newcommand{\sherpa}{S\protect\scalebox{0.8}{HERPA}\xspace}
\newcommand{\pythia}{P\protect\scalebox{0.8}{YTHIA}\xspace}

\newcommand{\rivet}{R\protect\scalebox{0.8}{IVET}\xspace}
\newcommand{\fastjet}{F\protect\scalebox{0.8}{AST}J\protect\scalebox{0.8}{ET}\xspace}

\newcommand{\qg}{\emph{q/g}}

\usepackage{cprotect}

\begin{document}

\title{Tagging the initial-state gluon}
\author{
Simone Caletti\thanksref{e1,addr1}
\and
Oleh Fedkevych\thanksref{e2,addr1}
\and
Simone Marzani\thanksref{e3,addr1}
\and
Daniel Reichelt\thanksref{e4,addr2}
}

\thankstext{e1}{e-mail: simone.caletti@ge.infn.it}
\thankstext{e2}{e-mail: oleh.fedkevych@ge.infn.it}
\thankstext{e3}{e-mail: simone.marzani@ge.infn.it}
\thankstext{e4}{e-mail: daniel.reichelt@uni-goettingen.de}

\institute{Dipartimento di Fisica, Universit\`a di Genova and INFN, Sezione di Genova, Via Dodecaneso 33, 16146, Italy\label{addr1}
\and
Institut f{\"u}r Theoretische Physik, Georg-August-Universit{\"a}t G\"ottingen, Friedrich-Hund-Platz 1, 37077 G\"ottingen, Germany\label{addr2}
}

\date{}


\maketitle

\begin{abstract}
We study the production of an electroweak boson in association with jets, in processes where the jet with the highest transverse momentum is identified as quark-initiated. 
The quark/gluon tagging procedure is realised by a cut on a jet angularity and it is therefore theoretically well-defined and exhibits infrared and collinear safety. 
In this context, exploiting resummed perturbation theory, we are able to provide theoretical predictions for transverse momentum distributions at a well-defined and, in principle, systematically improvable accuracy. 
In particular, tagging the leading jet as quark-initiated allows us to enhance the initial-state gluon contribution.
Thus these novel transverse momentum distributions are potentially interesting observables to probe the gluonic degrees of freedom of the colliding protons. 
\end{abstract}


\section*{Introduction}

Over the past decade our understanding of the internal structure of jets has increased tremendously. Thanks to the applications of the methods of perturbative QCD, the field has become mature and jet substructure algorithms that are, at the same time, performant and robust have been developed. Observables originally designed for searching new physics are now the target of unfolded measurements that can be compared to theoretical predictions, at the precision level, see \textit{e.g.}~\citep{Aad:2020zbq,Aad:2019vyi,Aad:2019onw,Aaboud:2017qwh,Aaboud:2019aii,Sirunyan:2018xdh,Sirunyan:2018gct,CMS:2021vsp}. Furthermore, ideas developed by the jet substructure community have found applications in other contexts of particle physics. For instance, after it was realised that the momentum fraction that characterises the splitting identified by the \SD algorithm~\citep{Larkoski:2014wba} follows a distribution dictated by the QCD splitting functions~\citep{Larkoski:2015lea, Larkoski:2017bvj, Tripathee:2017ybi, Cal:2021fla}, this observable has become a standard way of probing interactions with the quark/gluon plasma, see e.g.~\citep{Chen:2021osv} and references therein. 

This work is part of an incipient effort to find innovative ways of applying jet substructure techniques to the broader LHC physics program and to provide the necessary tools so that this cross-pollination can bear its fruits.  In this context, measurements of the internal structure of highly energetic jets produced in association with a boosted electroweak boson offer an ideal playground for such studies, because the leptonic decay of the $Z$ boson offers a valuable trigger.
These events are not only relevant for background studies~\footnote{This process
  represents the main background for the associated production of a Higgs and an
  electroweak boson in the boosted regime, where the decay products of the Higgs
  are reconstructed in one jet.} but, as we shall detail in the following, they open up novel possibilities to employ jet substructure for Standard Model measurements. 

Substructure observables, such as jet angularities~\cite{Larkoski:2014pca}, measure the pattern of the had\-ronic activity within a jet. For this reason, they are often employed as tagging variables that aim to distinguish jets that have been originated by elementary particles carrying different colour degrees of freedom, \textit{e.g.} colour singlets versus QCD partons or, even, quarks versus gluons. 
Although there exist more powerful quark/gluon (\qg) discriminants than a cut on a jet angularity, this procedure is theoretically well-defined, infra-red and collinear (IRC) safe and, hence, its behaviour can be understood with perturbative methods.

In the following, we study $Z$+jet production, requiring that the jet with the highest transverse momentum has been identified as quark-initiated.  
As preliminarily explored in \citep{Amoroso:2020lgh}, by tagging the final-state, we indirectly bias particular partonic sub-processes. If the leading jet is tagged as quark-initiated, then, at leading order, the subprocess that features a quark and gluon in the initial state is enhanced. Thus, this procedure could potentially provide us with a new handle on the determination of the gluon parton distribution function (PDF).
The main result of this study is the calculation of the transverse momentum distribution of the $Z$ boson in events where the leading jet is identified as quark-initiated. Because of IRC safety of the tagging procedure, we are able to compute this distribution at a well-defined and, in principle improvable, accuracy. Therefore, this observable could be directly included in standard fits of PDFs.
The results presented here account for the resummation of the tagging parameter at the \emph{next-to-leading logarithmic} (NLL) accuracy, matched to fixed order predictions at $O\left(\alpha_s^2 \right)$ with respect to the Born contribution, henceforth denoted as next-to-leading order (NLO)~\footnote{The fixed-order counting is slightly different than what is usually employed in standard transverse-momentum distribution. This has to do with the fact that, in order to obtain a non-vanishing value of any angularity, the jet must have at least two constituents. Thus, in what follows we will refer to the lowest-order ($2 \to 2$) scattering as Born approximation, while (N)LO will be reserved for the contributions with one (two) additional emission(s).}.

\begin{figure*} 
\includegraphics[width=0.33\textwidth]{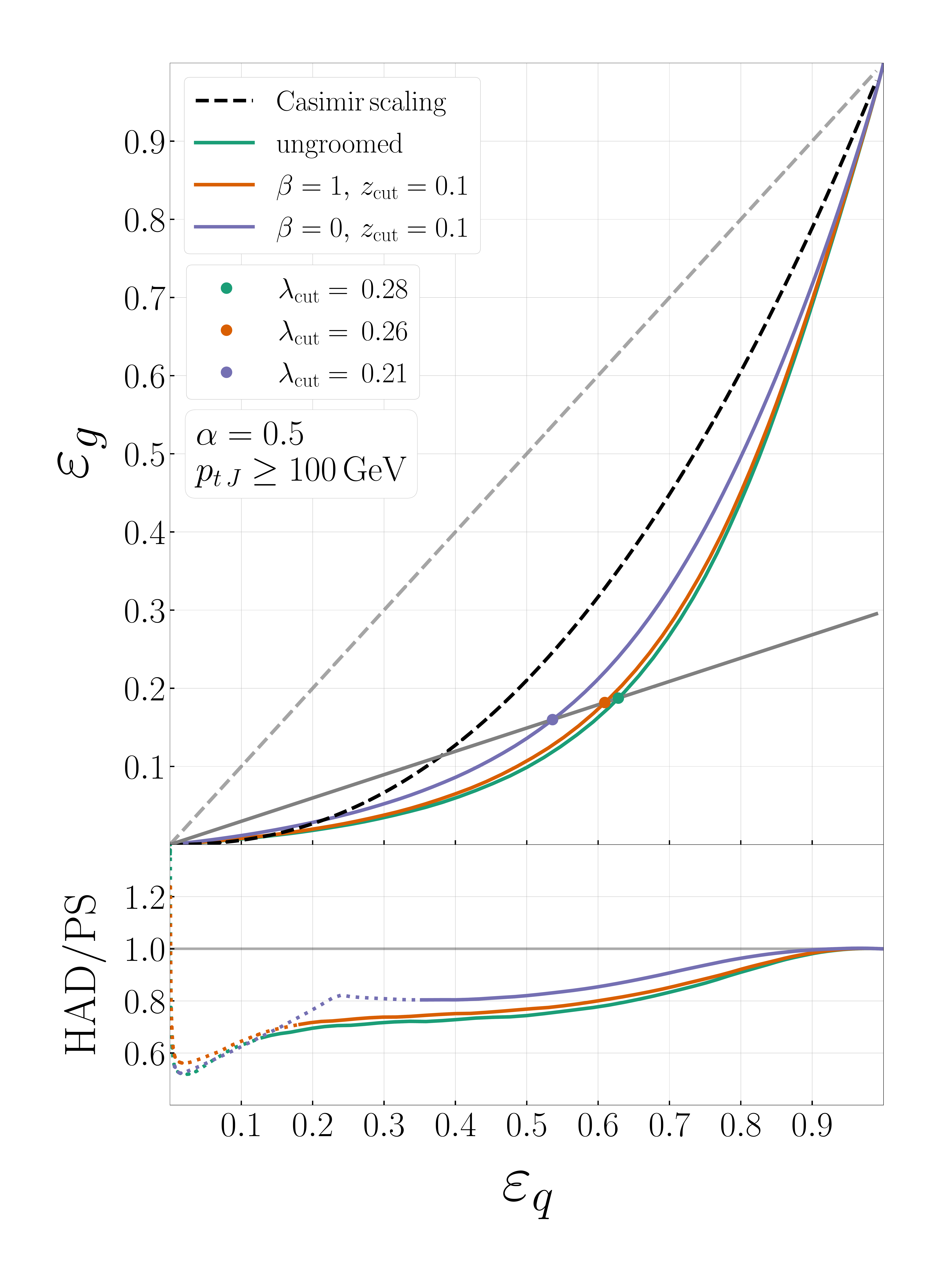}%
\hfill%
\includegraphics[width=0.33\textwidth]{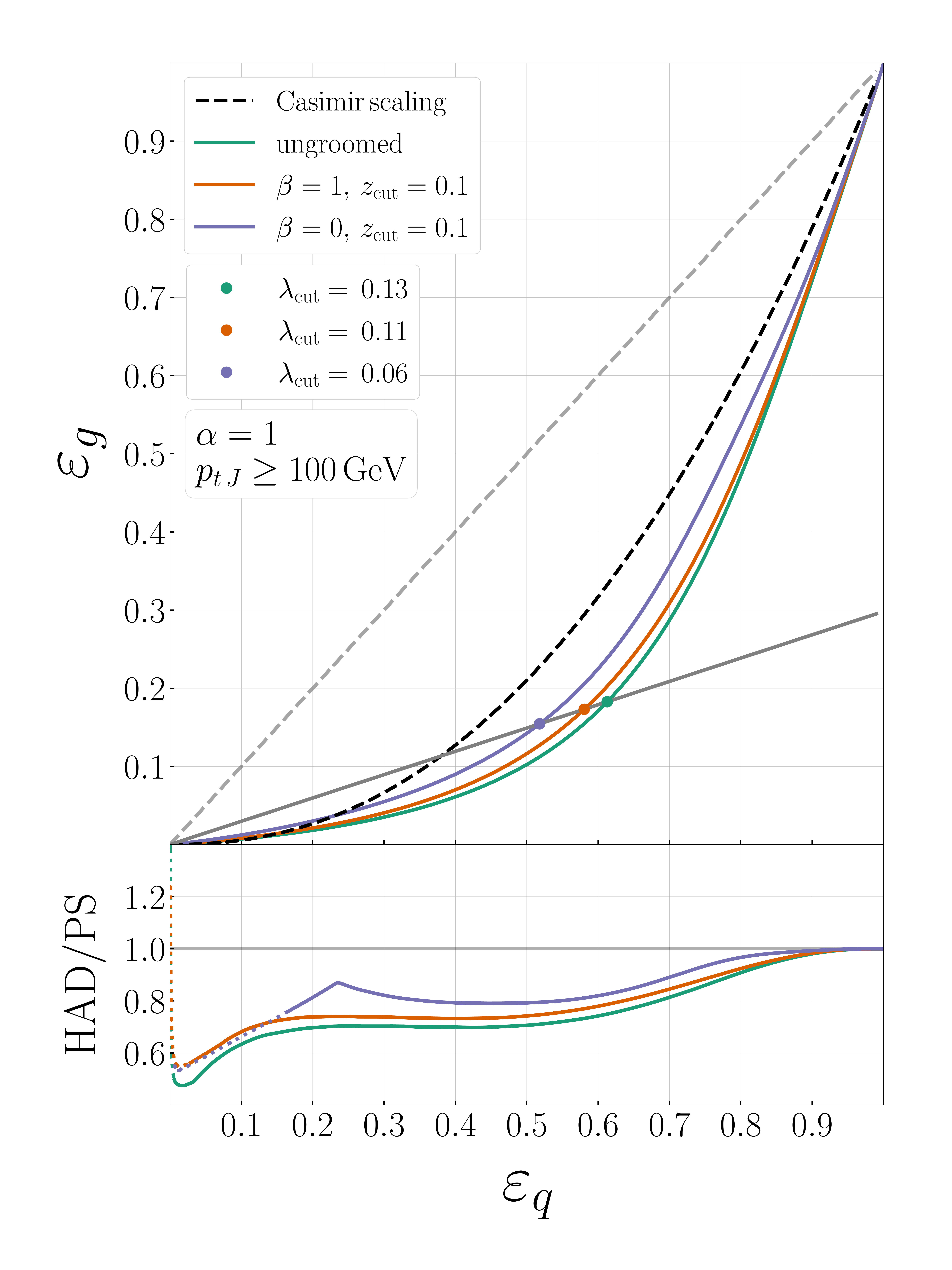}
\hfill%
\includegraphics[width=0.33\textwidth]{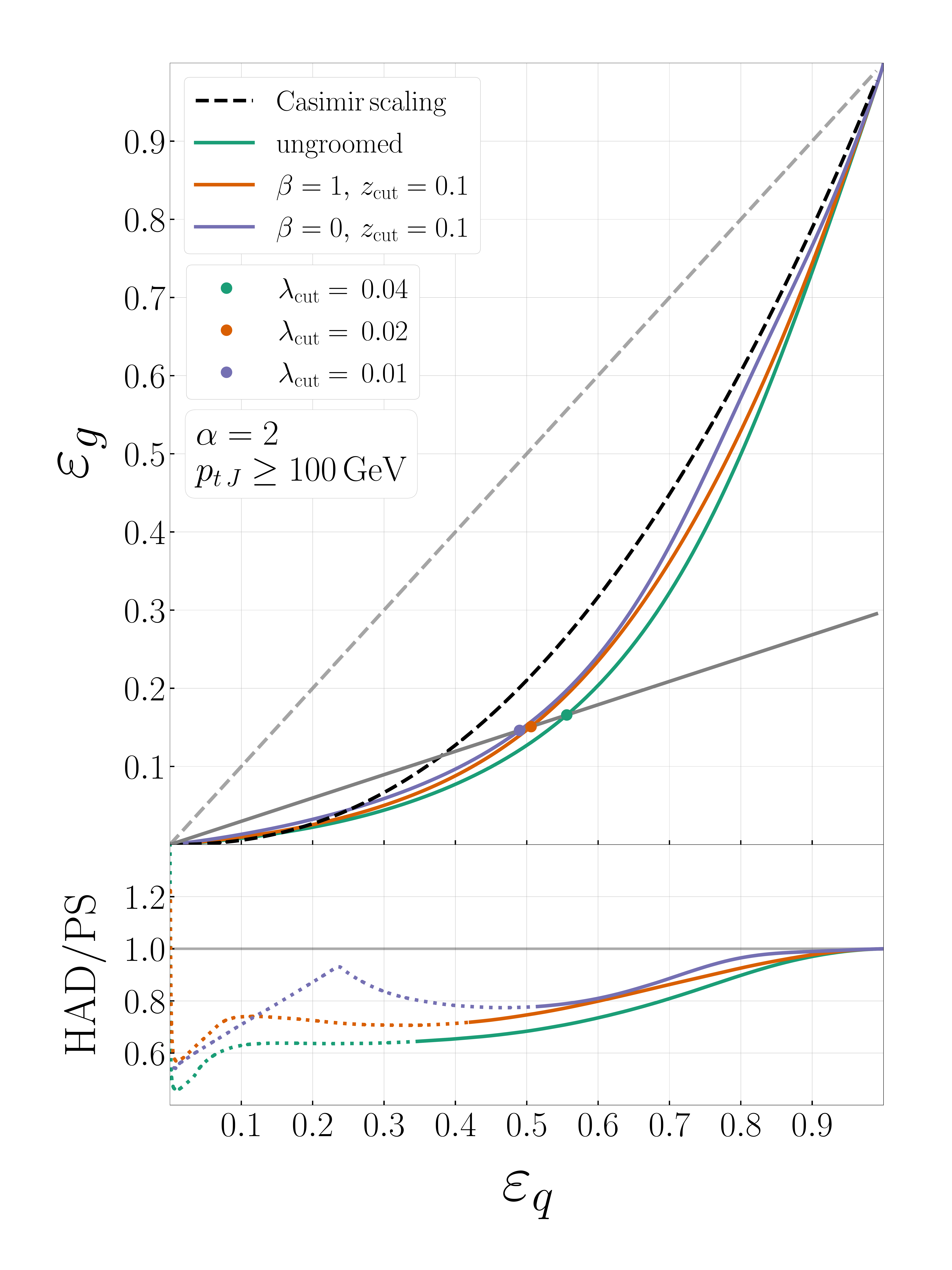}
\caption{Receiver Operating Characteristic (ROC) curves for different values of the angularity exponent $\alpha=0.5,1,2$, from left to right obtained with MC simulation using \pythia. In each plot, different curves refer to ungroomed jets and \SD jets with $\zc=0.1$ and $\beta=0,1$. 
For comparison we also show the ROC curves corresponding to a random classifier (dashed grey) and the CS ones (dashed black).
Straight lines represent the target performance, as detailed in the text. 
At the bottom of each plot we show the ratio between each ROC curve obtained at hadron-level and the corresponding parton-level result. 
The dashed portion of these lines indicate the region sensitive to splittings with relative transverse momentum below 1~GeV, where NP effects are expected to be sizeable. This region of phase-space is sensitive to shower cut-off effects, which cause the observed kinks.
} 
\label{fig:roc-curves}
\end{figure*}

\section*{Enhancing the gluon contribution}

We start by considering the channel fractions $f_{ij}$ that measure, to a given order in perturbation theory, the contribution to the $Z$ (or jet) transverse momentum distributions from the subprocess initiated by partons $i$ and $j$, which, for brevity, we indicate as $\sigma_{ij}^a$, where $a$ could be the $Z$ boson or the leading jet $J$:
\begin{equation}\label{frac-before-tagging}
f_{ij}^a=\frac{\sigma_{ij}^a}{\sigma^a_{qq}+\sigma^a_{q g}+\sigma_{gg}^a}, \quad a=Z, J,
\end{equation}
where $qq$ and $q g$ include any combination of quarks and anti-quarks. Our aim is to study how tagging a quark-jet in the final-state changes the relative contributions of the various partonic subprocesses. To this purpose, we define 
\begin{equation}\label{frac-after-tagging}
\widetilde{f}_{ij}^a=\frac{\widetilde{\sigma}_{ij}^a}{\widetilde{\sigma}^a_{qq}+\widetilde{\sigma}^a_{q g}+\widetilde{\sigma}_{gg}^a}, \quad a=Z, J,
\end{equation}
where the tilde indicates ``after tagging''. Furthermore, in what follows we will be interested in the fractions of the event (before and after tagging) which feature at least one gluon in the initial state. To this purpose, we define
\begin{equation} \label{gluon-f-before-after}
f_g^a=  f_{qg}^a +f_{gg}^a, \quad \text{and} \quad   \widetilde{f}_g^a=  \widetilde{f}_{qg}^a +\widetilde{f}_{gg}^a.
\end{equation}
We will refer to this fraction as \emph{gluon channel purity}.

We begin our discussion in a simplified setting, which is nevertheless enough to capture the essential physics points. We will then validate our conclusions using Monte Carlo (MC) parton shower simulations.
If we consider the Born approximation, then we have no $gg$ contribution and, because the $Z$ boson and the jet are back-to-back, we obtain $f_{ij}^Z=f_{ij}^J$.
Fragmentation of the hard partons can lead to transverse momentum imbalance, however, we can limit ourselves to the  leading logarithmic (LL) regime, where all parton splittings happen in the soft and collinear limit, with no recoil.~\footnote{We will lift this approximation when dealing with actual simulations. However, we can always choose a set of kinematical cuts that favours the back-to-back configuration.}
Therefore, dropping the superscript $a$, we have
\begin{equation}\label{frac-gluon-before-tagging}
f_g =\frac{\sigma_{qg}}{\sigma_{qq}+\sigma_{q g}}.
\end{equation}
We note that, within our approximation, the fraction of events with a (properly defined) final-state quark can be considered as a proxy for the gluon channel purity.
We find that, for $\ptj \ge 100$~GeV, $f_g \simeq 0.85$ and it exhibits a rather mild dependence on the transverse momentum. Next, we note that in our approximation we simply have
\begin{align}\label{frac-gluon-after-tagging}
\widetilde{f}_{g}&=\frac{\varepsilon_q  \sigma_{qg}}{\varepsilon_g \sigma_{qq}+ \varepsilon_q \sigma_{q g}}
=\left(1+ \frac{1-f_g}{f_g}\frac{\varepsilon_g}{\varepsilon_q} \right)^{-1},
\end{align}
where $\varepsilon_q$ is the efficiency of the tagger to correctly label quark jets and $\varepsilon_g$ the false-positive rate, which measures how often gluon jets are wrongly labelled as quarks. 
We note that a perfect quark-tagger with $\varepsilon_q=1$ and $\varepsilon_g=0$
returns tagged events that have been originated by one gluon
in the initial state, i.e.\ $\widetilde{f}_{g}=1$. On the other hand, with an
efficiency of 50\%, which corresponds to tossing a coin, we recover
Eq.~(\ref{frac-gluon-before-tagging}). 

A rather common class of \qg ~taggers exhibits at LL a property known as Casimir scaling (CS)~\citep{Larkoski:2014pca}, namely the tagging efficiencies are related by
$
\varepsilon_g = \left( \varepsilon_q \right)^{C_A/C_F},
$
where  the exponent is given by the ratio of the Casimir operators in the appropriate colour representation, $C_F$ for quarks and $C_A$ for gluons. 
This property emerges because the LL distributions of both quarks and gluons exhibit the same Sudakov-like behaviour, with a coefficient determined by the appropriate colour factor. 
Thus, for a CS tagger, we find
\begin{equation}\label{frac-gluon-after-tagging-CS}
\widetilde{f}_{g}^\text{CS}=\left(1+ \frac{1-f_g}{f_g}{\varepsilon_q }^{\frac{C_A}{C_F}-1} \right)^{-1}.
\end{equation}
For instance, a CS tagger with $\varepsilon_q=0.65$, yields $\widetilde{f}_{g}^a\simeq0.9$. This can be pushed to 0.95 if the tighter working point $\varepsilon_q=0.35$ is considered.
Taggers that obey CS are not the most performant, but they are interesting for us because they are under good theoretical control. 
The efficiencies $\varepsilon_i$ can be computed in QCD using resummed perturbation theory to a well-defined and, in principle improvable, theoretical accuracy. 
Furthermore, the inclusion of higher logarithmic corrections generally leads to an improvement with respect to strict CS, so that, depending on the specifics of the tagging procedure, the target gluon channel purity $\widetilde{f}_{g}^a\simeq0.95$ can be achieved at a reasonable working point. 
In the following we construct a CS tagger that is based on a particular class of jet substructure observables known as jet angularities.

\section*{Jet angularities as a quark/gluon tagger}
Jet angularities~\citep{Larkoski:2014pca} are defined as
\begin{equation}\label{eq:ang-def}
\lambda_\alpha= \sum_{i \in \text{jet}} \frac{p_{t,i}}{\ptj} \left(\frac{\Delta_i}{R_0} \right)^\alpha, 
\end{equation}
where the sum runs over the constituents of the hardest jet in the event and $ \Delta_i=\sqrt{(y_i-y_J)^2+(\phi_i-\phi_J)^2}$ is the distance in the azimuth-rapidity plane of particle $i$ from the jet axis. 
We define jets with the anti-$k_t$ clustering algorithm~\citep{Cacciari:2008gp} with radius $R_0$ and standard $E$-scheme for
recombination.
IRC safety requires $\alpha>0$, while angularities with $\alpha \le 1$ are sensitive to recoil against soft emissions~\citep{Larkoski:2013eya}. 
In order to circumvent this issue, when $\alpha \le 1$, the jet axis is obtained using the Winner-Take-All (WTA) recombination scheme~\citep{Larkoski:2014uqa}.
We also consider groomed jets.  In this case,
we recluster the jet with the Cambridge-Aachen algorithm~\citep{Dokshitzer:1997in, Wobisch:1998wt} and apply
the \SD grooming algorithm with parameters $\zc$ and
$\beta$~\citep{Larkoski:2014wba}. The angularity is then
computed on the constituents of the groomed jet, with the WTA prescription adopted for angularities with $\alpha \le 1$.

Because of the different colour factor, angularity distributions for quark- and gluon-initiated jets peak at different values. We can exploit this separation and define our \qg ~tagger through a cut on the jet angularity. In particular, a jet with $\lambda_\alpha<\lambda_\text{cut}$ will be labelled as a quark jet:
\begin{equation}
\widetilde{\sigma}_{ij}^a= \int_0^{\lambda_\text{cut}}d \lambda_\alpha \frac{d {\sigma}_{ij}^a}{d \lambda_\alpha},
\end{equation}
where we have introduced the differential angularity distribution. The implicit dependence of the tagged distribution on the transverse momentum, on the angularity exponent $\alpha$ and, optionally, on the \SD parameters $\zc$ and $\beta$, is understood. 

From the simple CS analysis above, we have concluded that we should work with quark efficiencies $\varepsilon_q\simeq 0.35$, in order to reach a purity of initial-state gluons around 0.95. The value  of the angularity cut that is necessary to achieve this working point for the tagger clearly depends on the angular exponent $\alpha$ in Eq.~(\ref{eq:ang-def}) as well as on the parameters $\zc$ and $\beta$ of the \SD algorithm, should we wish to employ groomed jets.
Different theoretical considerations can guide us with this choice. First of
all, we would like to preserve calculability, \textit{i.e.}\ we want to cement
our findings in perturbative field theory. Thus, we would like our tagger to be
as insensitive as possible to non-perturbative (NP) contributions such as
hadronisation corrections and the Underlying Event (UE). Secondly, although we
can calculate transverse momentum spectra in resummed perturbation
theory~\citep{Kang:2018qra,Kang:2018vgn,Caletti:2021oor}, we aim for
perturbative stability. Thus, we favour working points for the tagger for which
$\lambda_\text{cut}$ is not too small.

\begin{figure}
\begin{center}
\includegraphics[width=0.4\textwidth]{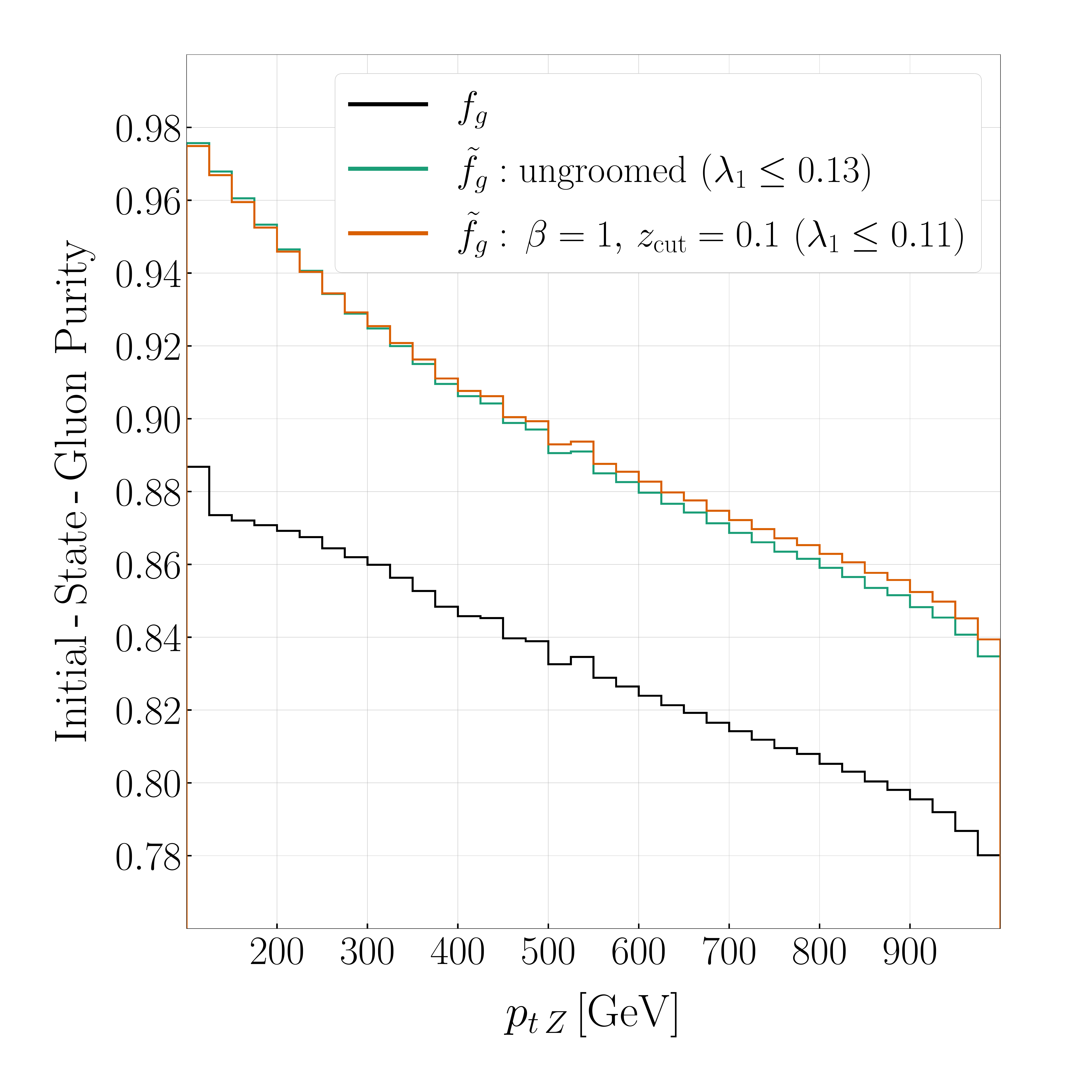}
\cprotect\caption{The initial-state gluon purity before ($f_g$) and after tagging ($\widetilde{f}_g$) obtained with \pythia simulations, as a function of the transverse momentum of the $Z$ boson.
}
\label{fig:initial-state-fractions}
\end{center}
\end{figure}

In order to turn the above considerations into a quantitative study we use simulated data obtained with the MC  event generator \pythia~8.303 \citep{Sjostrand:2014zea}. 
The UE is simulated according to  the model presented
in~\citep{Sjostrand:1985vv, Sjostrand:1987su,Sjostrand:2004pf} and hadronisation
effects according to the Lund string model~\citep{Andersson:1983ia,
  Sjostrand:1984ic}. 
Throughout the paper, we use the NNPDF 3.0 NLO set of PDFs \citep{NNPDF:2014otw}.
We consider the inclusive production of a pair of oppositely charged muons in proton--proton collisions at
$13~\text{TeV}$ centre-of-mass energy, requiring that the invariant mass of the muon pair to be within 70 and 110 GeV. 
Jets are clustered with the anti-$k_t$ algorithm with $R_0=0.4$ and ordered in transverse momentum. Henceforth, the jet will be implicitly considered to be the hardest one and we will refer to the muon-antimuon pair as the $Z$ boson. With this in mind, the fiducial volume is defined following Ref.~\citep{CMS:2021vsp}:
$p_{t\,\mu} > 26~\mathrm{GeV}$,  $\ptZ>30\;\text{GeV}$, $ \ptj > 15~\mathrm{GeV}$, $ |\eta_\mu|<2.4$, and $ |y_{\text{jet}}| < 1.7$.
Furthermore, in order to enforce back-to-back configurations, we impose 
$
 \left| \frac{\ptj - \ptZ}{ \ptj + \ptZ}  \right| < 0.3$,
 $
\left|\phi_{\rm jet}-\phi_Z \right| > 2.
$

For event selection and analysis we employ \rivet~\citep{Buckley:2010ar,Bierlich:2019rhm}. Jet reconstruction is done with
 \fastjet~\citep{Cacciari:2011ma}, and the \SD implementation in the \fastjet ~\verb|contrib| is used.
We distinguish two different stages of the simulation: ``parton-level'',
\textit{i.e.}\ with parton shower effects only, and ``hadron-level'', \textit{i.e.} with
hadronisation and UE included.
By  keeping the two partonic processes of interest separate, we compute Receiver Operating Characteristic (ROC) curves that show the mis-tag rate (gluon efficiency) $\varepsilon_g$ as a function of the signal (quark) efficiency. They are computed for different values of the angularity exponent $\alpha=0.5, 1, 2$ in the ungroomed case and for \SD jets with $\zc=0.1$ and $\beta=0,1$. We show hadron-level results in Fig.~\ref{fig:roc-curves}, as well as the ratios to their parton-level counterparts, which we take as a measure of NP contributions. 
The dotted portion of the latter indicates that the efficiency $\varepsilon_q$ is dominated by splittings in the non-perturbative region, as determined, for instance, in Ref.~\cite{Caletti:2021oor}.
We also show the target line, which fixes the gluon efficiency $\varepsilon_g$ as a function of $\varepsilon_q$, for given $f_g$ and $\widetilde{f}_g$,
\begin{equation}
\varepsilon_g= \frac{f_g (1-\widetilde{f}_g)}{\widetilde{f}_g(1-f_g)}\varepsilon_q,
\end{equation} 
which is easily derived from Eq.~(\ref{frac-gluon-after-tagging}).
The slope shown in Fig.~\ref{fig:roc-curves} is determined by the original $f_g=0.85$ and target gluon purity $\widetilde{f}_g=0.95$. The intersections of each ROC curve with the target line set our tagger working points. Corresponding values of $\lambda_\text{cut}$ are reported in the figure.
The choice of the tagger working points is clearly not unique. For instance, we could have optimised the signal-to-background ratio by choosing on each ROC curve the point that is closest to the $(1,0)$ corner. 

The analysis of Ref.~\citep{Caletti:2021oor} tells us that larger values of $\alpha$ are under better theoretical control. However, as it is clear from Fig.~\ref{fig:roc-curves}, lower values of $\alpha$ have better performance, essentially because of their increased sensitivity to the collinear region.
Thus, the choice $\alpha=1$ appears to be a good compromise between performance and robustness. 
We note that the use of \SD does not always provide us with improvements on the
size of the NP contributions.  
This might be related to the fact that in order to obtain the same efficiency
$\varepsilon_q$ we need to cut groomed jets at lower values of
$\lambda_\text{cut}$, where NP physics may be more prominent. 
Furthermore, we note that we are working with a rather small jet radius, which prevents large contributions from the UE. We expect (light) grooming to be beneficial, should one consider larger jet radii. 
Finally, configurations with aggressive \SD, i.e.\ $\beta=0$, typically result in worse performance, because important information is groomed away. 
Thus, in what follows, we shall focus on the $\alpha=1$ case either with no grooming or with 
$\beta=1$, $\zcut=0.1$.

We now quantify the gain in the initial-state gluon purity that we obtain after tagging. Fig.~\ref{fig:initial-state-fractions} shows the fractions $\tilde{f}_g^Z$ as a function of the $Z$ boson transverse momentum for the selected taggers, with $f_g^Z$ also shown for comparison. 
We first note that the performance of both taggers is very good, leading to gluon channel purities that exceed our 95\% target.
Analogous conclusions can also be drawn if we plot our results as a function of $\ptj$. However, as we will argue shortly, in this context, the transverse momentum distribution of the $Z$ boson is a more robust observable. 
We also notice that the gluon channel purities decrease with $\ptZ$. This is due to the fact that our taggers are defined looking at their efficiencies with $\ptj>100$~GeV and, therefore, we expect them to work better at the lower end of the transverse momentum spectrum. This loss in performance could be cured by adjusting the cut on the angularity as a function of $\ptj$. However, in this first study, we prefer to keep our framework simple. 

\begin{figure}
\begin{center}
\includegraphics[width=0.4\textwidth]{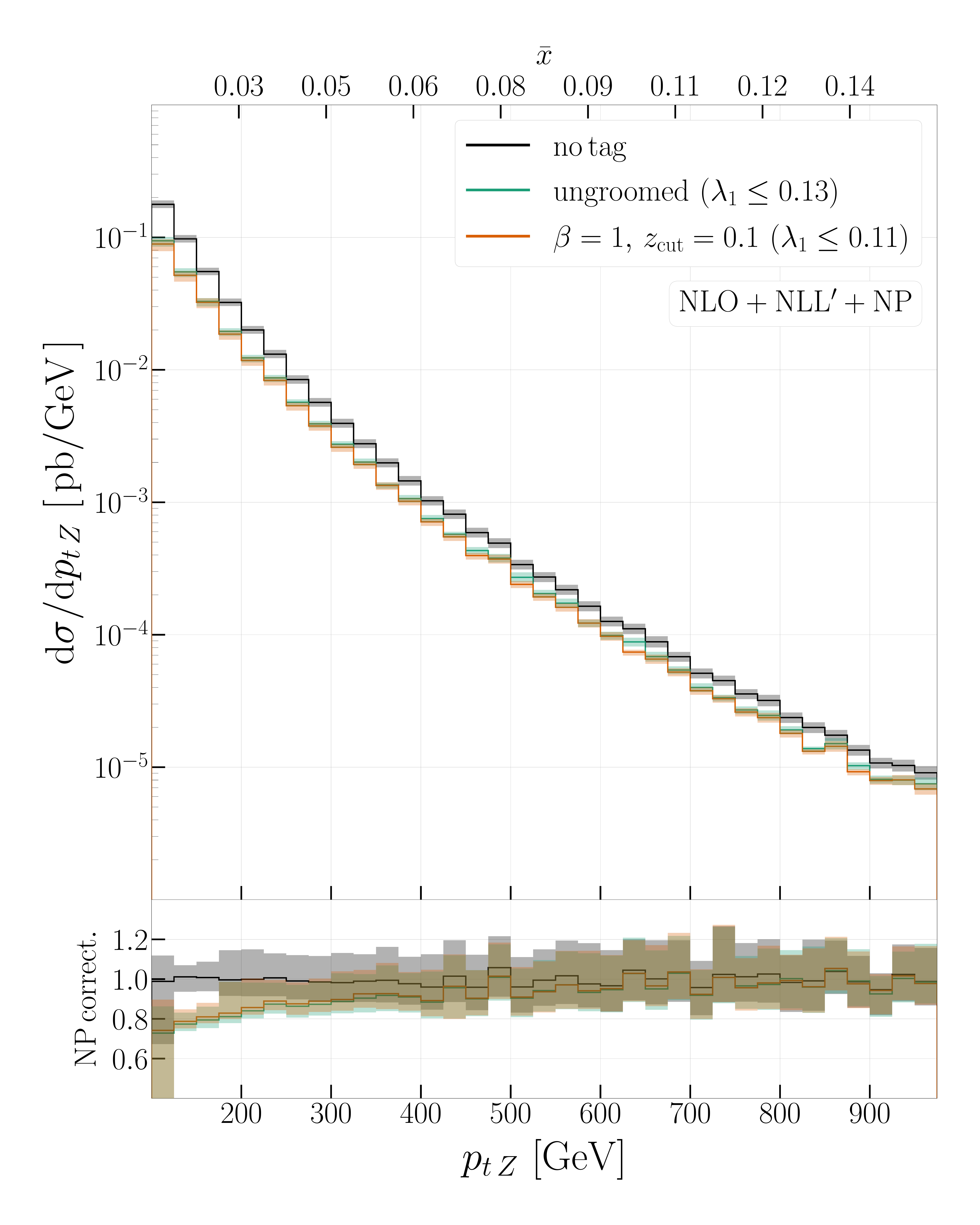}
\caption{Transverse momentum distribution of the $Z$ boson in $Z$+jet events, with the leading jet tagged as quark-initiated, according to our operational definition, detailed in the text. 
The untagged distribution is also shown for comparison. The NLO+NLL$'$ calculation is supplemented with a NP correction factor, shown at the bottom.}
\label{fig:pt-distributions}
\end{center}
\end{figure}

\section*{Transverse momentum distributions}
We now provide theoretical predictions for our observables of interest, namely transverse-momentum distributions, in the presence of tagging. 
Our calculation includes the resummation of logarithms of $\lambda_\text{cut}$
at NLL accuracy and is matched to NLO. Thanks to a
flavour-dependent matching procedure, \emph{cf.} also \cite{Banfi:2006hf, Baron:2020xoi}, we are able to achieve NLO+NLL$'$ accuracy.
We also include a bin-by-bin  NP  correction factor obtained with MC simulations. 
~\footnote{See Ref.~\cite{Caletti:2021oor} for details about the calculation and its numerical implementation in the resummation plugin
  \cite{Gerwick:2014gya, Baberuxki:2019ifp} to the \sherpa
  \cite{Gleisberg:2008ta, Sherpa:2019gpd} framework, including perturbative uncertainties, obtained by varying the perturbative (renormalisation, factorisation and resummation) scales  and
  NP corrections. We use COMIX \cite{Gleisberg:2008fv} in conjunction with OpenLoops \cite{Buccioni:2019sur} and
  Recola \cite{Actis:2016mpe, Biedermann:2017yoi} for the fixed order
  calculation. The NP corrections are based on \sherpa parton shower simulations
  at MC@NLO accuracy \cite{Frixione:2002ik, Hoeche:2011fd} hadronised with \sherpa's
  cluster fragmentation model \cite{Winter:2003tt}.}

Our results are reported in Fig.~\ref{fig:pt-distributions}, where we show the $\ptZ$ distribution for events where the highest-$p_t$ jet is quark initiated, for the taggers selected for this study.
We show the $\ptZ$ distribution with no-tagging, with tagging on standard jets and with tagging on \SD jets.
We note that NP corrections are rather sizeable at low $\ptZ$, making this observable most reliable in the high transverse momentum region. 
The latter is actually \emph{per se} interesting because it allows us to probe the
proton dynamics described by the PDFs at large values of the momentum fraction
$x$, \emph{i.e.} in a kinematic region where they are less constrained. To
qualitatively assess the values of $x$ which would be accessible, we
show, on the upper horizontal axis, the Born-level momentum fraction computed at
central (zero) rapidities:
$
\bar x=x_{1,2} = \frac{p_t e^{\pm y_J}+\sqrt{p_t^2+m_Z^2}e^{\pm y_Z}}{\sqrt{S}}\Big|_{y_J=y_Z=0}.
$

We have also studied $\ptj$ distributions. However, as anticipated, these distributions turn out to be less robust, essentially because 
jet dynamics can be significantly altered by the cut on the angularity, as well as by the grooming procedure~\footnote{Note that if $\beta=0$, the groomed $\ptj$ distribution is not even IRC safe.}. In contrast, the $\ptZ$ spectrum is inclusive with respect to the jet activity and thus, for a given working point of the tagger, less dependent on the details of the tagging procedure itself. We can therefore take modifications in such distribution as more directly related to the bias that the tagger induces on the composition of the partonic initial state, which is what we want to achieve.

\section*{Conclusions and future developments}
We have shown how \qg ~tagging can be successfully applied to $Z$+jets events in order to significantly enhance the gluon-initiated contributions. 
In particular, our tagger is realised through a simple cut on a jet angularity, which is an IRC safe observable and therefore can be studied in perturbation theory. Exploiting  MC simulations, we have performed a study of their efficiencies and their dependence on NP effects, exploring different angularities and different levels of grooming.
We have explicitly studied the transverse momentum of the $Z$ boson, providing theoretical predictions that included both all-order resummation and matching to NLO. We have shown that we can achieve initial-state gluon purities close to 95\%.

We see several possible directions for future work in this context. 
First, we would like to assess the impact of this type of observables on PDF fits.
In particular, while our study does show an increase in the gluon purity, this comes at the cost of a noticeable reduction of the available dataset. Thus, despite the relative large cross-section of the process we are considering, we may need to optimise the tagger's working point, taking this further aspect into consideration. For instance, as mentioned above, we could choose $\lambda_\text{cut}$ so that the signal-to-back\-ground ratio is maximised.
Second, the results presented in this study are based on a calculation of the angularity spectra, and hence of closely related efficiencies, at NLO+NLL$'$. The resummed calculation can be promoted to higher accuracy~\citep{Frye:2016okc,Frye:2016aiz,Kardos:2020gty}.
The fixed-order can also be improved by including the two-loop correction to $Z$+1\,jet (see~\citep{Gehrmann-DeRidder:2015wbt} and references therein). An improvement in the description of the angularity distribution away from $\lambda_\alpha=0$  is much more challenging
because it requires $Z$+2\,partons at NNLO, which may become available in the near future.
%
Finally, with the aim of enhancing performance while maintaining calculability, it would be interesting to consider more powerful, albeit more sophisticated, taggers. 
In this context, the Les Houches multiplicity~\citep{Amoroso:2020lgh} 
is very promising and, although its theoretical understanding is only in its infancy,  we believe that achieving NLL accuracy is within reach.

\section*{Acknowledgments}
We thank S. Schumann and G. Soyez for collaboration on related topics. 
We also thank our ATLAS and CMS colleagues: R.~Aggleton, J.~Ferrando, A.~Hinzmann,  M.~LeBlanc, B.~Nachman, and F.~Sforza, for useful discussions.
This work is supported by Universit\`a di Genova under the curiosity-driven grant ``Using jets to challenge the Standard Model of particle physics'' and by the Italian Ministry of Research (MUR) under grant PRIN 20172LNEEZ.
DR further acknowledges funding
from the European Union Horizon 2020 research and innovation programme as part
of the Marie Sklo\-dowska Curie Innovative Training  Network MCnet ITN3 (grant
agreement no. 722104), from BMBF (contract \\ 05H18MGCA1) and by the Deutsche
Forschungsgemeinschaft (DFG, German Research Foundation) - project number 456104544.
Figures  were
created with the Matplotlib \citep{Hunter:2007ouj} and NumPy \citep{NumPy} libraries.  

\bibliographystyle{jhep}       

\bibliography{references}

\end{document}